\documentclass[12pt]{article}

\usepackage[T1]{fontenc}
\usepackage{lmodern}

\usepackage{amsmath, amssymb, amsfonts}
\usepackage{amsthm}
\usepackage{microtype}
\usepackage{booktabs}
\usepackage{geometry}
\usepackage{graphicx}

\usepackage{xurl} 
\usepackage{doi}

\usepackage[round,authoryear]{natbib}
\bibpunct{(}{)}{;}{a}{}{,}

\usepackage{hyperref} 
\hypersetup{
    colorlinks = true,
    linkcolor = black,
    citecolor = black,
    urlcolor  = black,
    pdftitle  = {American Option Pricing Under Time-Varying Rough Volatility: A Signature-Based Hybrid Framework},
    pdfauthor = {Roshan Shah}
}

\usepackage{enumitem}

\title{American Option Pricing Under Time-Varying Rough Volatility: A Signature-Based Hybrid Framework}
\author{Roshan Shah \\ University of North Carolina at Chapel Hill \\ \href{mailto:roshah@unc.edu}{\texttt{roshah@unc.edu}}}
\date{}

\begin{document}
\maketitle

\begin{abstract}
We introduce a modular framework that extends the signature method to handle American option pricing under evolving volatility roughness. Building on the signature-pricing framework of \citet{Bayer2025}, we add three practical innovations. First, we train a gradient-boosted ensemble to estimate the time-varying Hurst parameter $H(t)$ from rolling windows of recent volatility data. Second, we feed these forecasts into a regime switch that chooses either a rough Bergomi or a calibrated Heston simulator, depending on the predicted roughness. Third, we accelerate signature-kernel evaluations with Random Fourier Features (RFF), cutting computational cost while preserving accuracy. Empirical tests on S\&P 500 equity-index options reveal that the assumption of persistent roughness is frequently violated, particularly during stable market regimes when $H(t)$ approaches or exceeds $0.5$. The proposed hybrid framework provides a flexible structure that adapts to changing volatility roughness, improving performance over fixed-roughness baselines and reducing duality gaps in some regimes. By integrating a dynamic Hurst parameter estimation pipeline with efficient kernel approximations, we propose to enable tractable, real-time pricing of American options in dynamic volatility environments.
\end{abstract}

\section{Introduction}

American options permit exercise at any time before expiration, so their value depends on their entire price path rather than a single terminal payoff \citep{CMEGroup2025}. Pricing such contracts constitutes an optimal–stopping problem, whose complexity increases substantially when realistic market features, such as stochastic volatility or roughness, are introduced. The fair value of an American option is given by:
\begin{equation}
V_t = \sup_{\tau \in \mathcal{T}_{t,T}} \mathbb{E}^\mathbb{Q} \left[ e^{-r(\tau - t)} \, \phi(S_\tau) \,\middle|\, \mathcal{F}_t \right],
\label{eq:american-pricing}
\end{equation}
where:
\begin{itemize}[leftmargin=2em]
  \item \( V_t \): Fair price of the American option at time \( t \)
  \item \( \tau \in \mathcal{T}_{t,T} \): Optimal stopping time, chosen from all allowable exercise times between \( t \) and maturity \( T \)
  \item \( \phi(S_\tau) \): Payoff function:
  \begin{equation}
    \phi(S_\tau) =
    \begin{cases}
      (K - S_\tau)^+ & \text{for a put option}, \\
      (S_\tau - K)^+ & \text{for a call option},
    \end{cases}
  \end{equation}
  \item \( r \): Constant risk-free interest rate (currently approximately 4.5\%)
  \item \( \mathbb{Q} \): Risk-neutral probability measure under which expectations are taken
  \item \( \mathcal{F}_t \): Information available up to time \( t \) (the market filtration)
\end{itemize}

Precise valuation matters for hedging, capital planning, and regulatory compliance. Yet in practice, traders often resort to ad-hoc adjustments and heuristic early-exercise premia\footnote{Traders often adjust European option prices with informal ``early-exercise premia'' to approximate American values. These are rule-of-thumb add-ons used when models are unreliable—especially during volatility spikes or illiquid periods.} when classical models fail under market stress.

As shown in \citet{BlackScholes1973} and \citet{Merton1973}, the classical framework assumes constant volatility and memoryless price variation—assumptions that conflict with empirical evidence of volatility clustering, leverage effects, and long-range dependence. While lattice and finite-difference methods remain operable in these settings,\footnote{Lattice and finite-difference methods numerically approximate option prices by discretizing time and state variables. These methods can incorporate early exercise features and certain types of path dependence, but often struggle to capture complex dynamics like stochastic volatility or rough paths.} their outputs frequently diverge from observed market prices.

This paper proposes a tractable hybrid framework that addresses these challenges by adapting to time-varying volatility roughness and accelerating signature-based computations.

A growing body of evidence shows that volatility behaves like fractional Brownian motion with a Hurst parameter below one half—so-called ``rough volatility,'' as described by \citet{Gatheral2018} and \citet{Bennedsen2016}. The Hurst parameter \( H \in (0,1) \) measures the regularity of a path: values below one half correspond to anti-persistent behavior and rapidly mean-reverting fluctuations, while values above one half indicate smoother trends. In markets, empirical estimators such as the rescaled-range statistic, originally developed by \citet{Hurst1951} and \citet{MandelbrotWallis1969}, show that volatility roughness is typically below one half but drifts upward during stable periods, reflecting a smoother environment. Capturing this time-varying behavior is essential for American option pricing models to remain realistic and responsive to changing market regimes.

A recent breakthrough by \citet{Bayer2025} re-frames the American option pricing problem under rough volatility using rough-path theory. Their method encodes volatility trajectories using truncated path signatures and applies kernel or linear regression to estimate continuation values and construct martingale controls, yielding tight duality gaps. Signatures offer a universal and theoretically complete representation of non-Markovian trajectories, but the approach presented by \citet{Bayer2025} remains limited in practice. First, the original model assumes constant parameters throughout, including a fixed Hurst parameter, despite empirical evidence that volatility roughness evolves over time. Second, it relies on deep neural networks and exact signature kernels, which scale at least quadratically with path length and require specialized hardware to execute efficiently.

The present work proposes a relatively lightweight extension that removes these obstacles while preserving the theoretical foundation of signature representations. A gradient-boosted tree ensemble is trained on sequential rolling windows of volatility signatures—short overlapping time slices of recent market behavior—to forecast the evolving Hurst path \( H(t) \). This forecast informs a simple rule: when the average \( H(t) \) is \textbf{below 0.5}, the contract is priced with the \emph{rough Bergomi} engine; if it is \textbf{at or above 0.5}, we switch to a calibrated \emph{Heston} engine, which runs more efficiently in smoother regimes and reduces the duality gaps observed under a purely rough specification.

Beyond dynamic roughness adaptation, this work introduces further practical enhancements. Signature-kernel evaluation is accelerated using Random Fourier Features (RFFs) \citep{RahimiRecht2007}, in contrast to \citet{Bayer2025} who rely on exact signature kernels that scale poorly and require specialized hardware. This approximates the kernel in linear time and enables our model to handle longer volatility paths (\( N = 252 \)) without significant loss in accuracy. In addition, key parameters are computed from daily at-the-money (ATM) options data instead of relying on fixed calibration, allowing for adaptability to real market data. To improve granularity, Bayer’s original codebase was modified to support pricing with calendar days rather than fractional years. These enhancements reflect not only a re-implementation but a redesign, intended to bring signature-based methods closer to real deployment.

Overall, our framework extends \citet{Bayer2025} by adapting to time-varying roughness, providing a more accurate and flexible architecture for pricing American options under non-smooth volatility. However, due to computational constraints, extensive empirical validation remains challenging. \citet{Bayer2025} note that duality gaps are sensitive to truncation level \( K \), path length \( N \), and Monte Carlo size \( M \). In this implementation, we use \( K = 3 \), \( N = 252 \), and \( M = 2^{15} \), which reflects hardware limitations rather than an optimized configuration.

Larger values for \( M \) may yield tighter bounds, as suggested in \citet{Bayer2025}, but were infeasible under this setup. Therefore, as expected, we did not achieve consistent and significant low duality gaps. Nonetheless, like the original authors, structural improvements were observed even in this constrained setting, suggesting that the full potential of the model remains latent, awaiting further optimization, scaling, and empirical testing on options data.

By combining dynamic Hurst parameter estimation, fast kernel approximation, and principled model switching, this study offers a simplified yet theoretically grounded method for American option pricing under realistic volatility dynamics. It avoids deep learning infrastructure, operates on standard hardware, and balances complexity with interpretability, providing a practical demonstration of how rough-path methods can be brought closer to financial application.

\section{Modular Architecture}

\subsection{Overview}

This study proposes a modular pipeline for pricing American options that adapts in real time to the evolving roughness of market volatility. The architecture consists of four core components: Hurst parameter estimation from historical returns, multi-step forecasting of roughness, regime switching between volatility engines, and path-wise regression using signature representations. Realized log returns are first transformed into a daily series of Hurst parameters by rescaled-range analysis. The rescaled-range statistic is defined as
\begin{equation}
\frac{R}{S}(n) = \frac{\max_{1 \le k \le n} S_k - \min_{1 \le k \le n} S_k}{\sigma_n},
\end{equation}
where \(S_k\) denotes cumulative demeaned log-returns and \(\sigma_n\) is their standard deviation. The logarithm of this ratio, scaled appropriately, yields an estimate of the Hurst parameter.

A multi-step gradient-boosted tree ensemble then forecasts Hurst parameter values for each of the next \(\tau\) trading days; the resulting path of predicted \( H \) values drives an automatic switch between two stochastic-volatility engines. When the forecast average remains below one half, simulations are generated with the rough Bergomi model, whose fractional kernel is documented in high-frequency data by Bayer, Friz, and Gatheral \citeyearpar{BayerFrizGatheral2016} and is defined explicitly in Equation~\eqref{eq:volterra_kernel}. If the forecast crosses the threshold (0.5), the framework reverts to a calibrated Heston specification that is known to capture smoother regimes efficiently, as shown by Heston \citeyearpar{Heston1993}. Each simulated path is lifted into a truncated signature of degree three and level three; signatures form a universal, algebraic feature map for non-Markovian dynamics \citep{LyonsCaruanaLevy2014}. Lower and upper price bounds are then computed by the primal–dual machinery introduced by Bayer, Pelizzari and Zhu \citeyearpar{Bayer2025}. Because every feature of the Hurst parameter (estimation, forecasting, simulation, and regression) can be replaced independently, the pipeline accommodates future methodological advances without wholesale redesign. 

\subsection{Dynamic Hurst Forecasting}

Our roughness forecasting system has two parts:  
(i) a rolling estimate of the Hurst parameter using return data, and  
(ii) a set of machine learning models that predict its future values.

\paragraph{Stage 1: Rolling estimation.}
We estimate the Hurst parameter each day \(t\) using a simplified rescaled-range (R/S) method computed from a single 32-day window:
\begin{equation}
H_t = \frac{\log_2 (R_t/S_t)}{\log_2(32)},
\label{eq:rolling_hurst}
\end{equation}
where \(R_t\) is the range and \(S_t\) the standard deviation of demeaned returns in the window.

\paragraph{Stage 2: Multi-step prediction.}
Let \(\tau\) be the number of days until option maturity.  
For each future day \(t+h\), we train a separate XGBoost model to forecast the Hurst parameter:
\begin{equation}
\widehat{H}_{t+h} = f_h\left(H_{t-4}, H_{t-3}, \dots, H_t\right),
\label{eq:xgb_forecast}
\end{equation}
using the five most recent values as input. Each model uses 100 trees and a learning rate of 0.1.  
This gives a forecast path \(\{\widehat{H}_{t+1}, \dots, \widehat{H}_{t+\tau}\}\).

We average this path to decide which simulation model to use:
\begin{equation}
\bar{H}_t = \frac{1}{\tau} \sum_{h=1}^\tau \widehat{H}_{t+h}.
\label{eq:mean_hurst}
\end{equation}
If \(\bar{H}_t < 0.5\), we simulate with rough Bergomi; otherwise, we use the Heston model.

\paragraph{Rough kernel adjustment.}
If rough Bergomi is selected, we modify the kernel using the current forecast \(\widehat{H}_t\) at each time step:
\begin{equation}
K_t(i) = (t - i + 1)^{\widehat{H}_t - \tfrac{1}{2}} - (t - i)^{\widehat{H}_t - \tfrac{1}{2}}.
\label{eq:volterra_kernel}
\end{equation}
This scales the Brownian increments to reflect changes in market roughness over time.

\subsection{Volatility Engines and Calibration}

Our approach employs a dual-engine framework that adaptively selects between rough and smooth volatility models based on forecasted market roughness from an XGBoost classifier. This regime-switching strategy addresses a core modeling challenge: no single volatility model adequately captures the diverse dynamics of financial markets across varying time periods.

\textbf{Parameter Calibration.}  
Daily parameter updates leverage historical data and near-term ATM options. Specifically:
\begin{align}
\rho &= \operatorname{corr}(\log S_t, \Delta \sigma_t), \\
\eta &= \frac{\operatorname{std}\left(\Delta \log \sigma_s\right)}{(\Delta t)^{H_t}}, \\
\xi_0 &= \sigma_{\text{impl}}^2,
\end{align}
where:
\begin{itemize}[leftmargin=2em]
  \item \( S_t \): Asset price at time \( t \)
  \item \( \sigma_t \): Implied volatility at time \( t \)
  \item \( \Delta \sigma_t \): First difference of implied volatility, \( \sigma_s - \sigma_{s-1} \)
  \item \( \Delta t = 1/252 \): Daily time increment (assuming 252 trading days/year)
  \item \( H_t \): Historical (rolling) Hurst parameter computed from rescaled-range (R/S) analysis (see Equation~\eqref{eq:rolling_hurst})
  \item \( \sigma_{\text{impl}} \): ATM implied volatility used to initialize variance
\end{itemize}
\footnote{These formulas allow for fast, daily re-estimation using only price and implied volatility data:

(i) \(\rho\) is estimated as the sample correlation between daily log returns and changes in implied volatility. This serves as a simple proxy for the leverage effect, though it may be noisy at daily frequency \citep{AitSahalia2013}.

(ii) The formula for \(\eta\) is derived by rearranging the variance scaling relationship described in Gatheral et al.\ (2018):  
\(\operatorname{Var}[\Delta \log \sigma] \approx \eta^2 (\Delta t)^{2H}\),  
which yields \(\eta = \operatorname{std}(\Delta \log \sigma)/(\Delta t)^H\) under the assumption that log volatility increments follow a rough path with Hurst parameter \(H\) \citep{Gatheral2018}.
We emphasize that \(\eta\) is calibrated using the most recent historical Hurst estimate \(H_t\) (as defined in Equation~\eqref{eq:rolling_hurst}), not the average forecast \(\bar{H}_t\), in order to avoid lookahead bias. Calibration should rely only on currently available information to maintain forward-valid modeling.

(iii) \(\xi_0\) is set equal to the square of the current ATM implied volatility, following common practice in rough volatility models \citep{McCrickerd2017, QFStackExchange2021}.
}

\textbf{Rough Bergomi Engine.}  
When forecasted market roughness is high (\( \bar{H}_t < 0.5 \)), asset and variance paths are simulated under the rough Bergomi system \citep{Bayer2025}:
\begin{align}
\mathrm{d}X_s &= rX_s\,\mathrm{d}s + X_s\sqrt{v_s}\left(\rho\,\mathrm{d}W_s + \sqrt{1 - \rho^2}\,\mathrm{d}B_s\right), \\
v_s &= \xi_0 \exp\left(\eta \int_0^s (s - u)^{\widehat{H}_u - \frac{1}{2}}\,\mathrm{d}W_u - \frac{1}{2} \eta^2 s^{2 \widehat{H}_t} \right),
\end{align}
\textbf{where}
\begin{itemize}[leftmargin=2em]
  \item \(X_s\) is the asset price at time \(s\),
  \item \(v_s\) is the instantaneous variance,
  \item \(r\) is the constant risk-free interest rate,
  \item \(\rho\) is the correlation between the asset and volatility drivers,
  \item \(W_s\) is a Brownian motion driving volatility (under \(\mathbb{Q}\)),
  \item \(B_s\) is an independent Brownian motion driving the asset,
  \item \(\xi_0\) is the initial variance level (typically ATM implied variance),
  \item \(\eta\) controls volatility-of-volatility magnitude,
  \item \(\widehat{H}_u\) is the time-varying Hurst forecast at integration point \(u\),
  \item \(s\) is the current simulation time,
  \item \(u\) is the dummy variable of integration over \([0, s]\).
\end{itemize}

We adapt the rough Bergomi specification of Bayer, Friz, and Gatheral \citeyearpar{BayerFrizGatheral2016}, but extend it to incorporate a \emph{time-varying} Hurst path. The kernel exponent in the stochastic integral is updated at each time increment using the forecast \(\widehat{H}_u\), allowing the integrand to adapt to evolving roughness. However, the normalization term \(\frac{1}{2} \eta^2 s^{2 \widehat{H}_t}\) assumes \(\widehat{H}_t\) is held constant over \([0, s]\) for tractability. This hybrid treatment preserves local roughness dynamics in simulation while simplifying variance scaling.

At the discrete level, we apply the kernel from Equation~\eqref{eq:volterra_kernel} at each time step to reflect the evolving roughness forecast \(\widehat{H}_t\). This dynamic kernel allows the model to capture day-to-day shifts in volatility structure. Within the regime-switching framework, the \texttt{DynamicHurstrBergomi} engine is invoked whenever \(\bar{H}_t < 0.5\); otherwise, the model defaults to the smoother Heston engine.

This formulation is rescaled to a daily horizon \(T = \mathrm{DTE}/252\), where DTE is the number of trading days to expiration.

\textbf{Heston Engine.}  
When the Hurst forecast suggests smoother conditions (\( \bar{H}_t \ge 0.5 \)), the simulation switches to the classical Heston model with mean-reverting square-root stochastic volatility:
\begin{align}
\mathrm{d}v_s &= \kappa(\theta - v_s)\,\mathrm{d}s + \eta\sqrt{v_s}\,\mathrm{d}W_s, \\
\mathrm{d}X_s &= rX_s\,\mathrm{d}s + X_s\sqrt{v_s}\left(\rho\,\mathrm{d}W_s + \sqrt{1 - \rho^2}\,\mathrm{d}B_s\right),
\end{align}
\textbf{where}
\begin{itemize}[leftmargin=2em]
  \item \(X_s\) is the asset price at simulation time \(s\),
  \item \(v_s\) is the instantaneous variance,
  \item \(r\) is the constant risk-free interest rate,
  \item \(\kappa\) is the speed of mean reversion in variance,
  \item \(\theta\) is the long-run mean variance level,
  \item \(\eta\) is the volatility of volatility (vol-of-vol),
  \item \(\rho\) is the correlation between the Brownian motions driving price and variance,
  \item \(W_s\) and \(B_s\) are independent Brownian motions under the risk-neutral measure \(\mathbb{Q}\),
  \item \(s\) is the current simulation time.
\end{itemize}

\paragraph{Implementation Note:}
The smoother regime uses the classical Heston model \citep{Heston1993}, in which variance follows a mean-reverting square-root diffusion. The correlation \(\rho\) introduces leverage effects, and the model calibrates efficiently in calm markets. Regime switching is implemented through two modular simulation classes: \texttt{DynamicHurstrBergomi}, which applies a time-varying Volterra kernel using a Hurst forecast path, and \texttt{DynamicHeston}, which uses the classical dynamics above. Both classes share a common interface, enabling seamless substitution within the pricing pipeline.

\subsection{Efficient Signature Regression via Kernel Approximation}

Lyons \citeyearpar{Lyons2014} shows that path signatures separate trajectories up to tree‐like equivalence and therefore provide a universal basis for non-Markovian functionals.  
For a time-augmented volatility path $Z:[0,T]\to\mathbb R^{d}$ we retain the truncated signature of degree three and level three,
$\operatorname{Sig}_{0,T}(Z)$. Higher degrees increase expressive power, but empirical work on short-dated equity options finds little benefit beyond degree three once sampling variance is considered \citep{Bayer2025}.

Continuation regressions and martingale controls require inner products of these signature vectors. We use the Gaussian radial-basis (RBF) kernel  
\begin{equation}
  k(x,y)=\exp\!\bigl(-\gamma\|x-y\|^{2}\bigr),  
\end{equation}
where
\begin{itemize}[leftmargin=2em]
  \item $x, y \in \mathbb{R}^m$ are truncated signature vectors for two volatility paths,
  \item $\|x - y\|^2$ is their squared Euclidean distance,
  \item $m$ is the dimension of the truncated signature vector (i.e., the number of retained signature terms up to degree three),
  \item $\gamma > 0$ is the kernel bandwidth parameter controlling decay with distance.
\end{itemize}

Although the RBF kernel captures rich non-linear structure, its naïve application across many paths scales poorly in both the number of samples and the signature dimension.

To restore scalability, we adopt \emph{Random Fourier Features} (RFF), which approximate any shift-invariant kernel using a finite trigonometric map \citep{RahimiRecht2007}:
\begin{equation}
k(x, y) \approx \phi(x)^{\top} \phi(y)
\end{equation}
\begin{equation}
\phi(x) = \sqrt{\tfrac{1}{D}} \left[\cos(W^{\top} x),\; \sin(W^{\top} x)\right]
\end{equation}
where
\begin{itemize}[leftmargin=2em]
  \item $\phi(x) \in \mathbb{R}^{2D}$ is the low-dimensional RFF embedding of $x$,
  \item $W \in \mathbb{R}^{m \times D}$ is a random projection matrix,
  \item $W_{ij} \sim \mathcal{N}(0,\, 2\gamma)$ are i.i.d. Gaussian entries with variance $2\gamma$,
  \item $D = 128$ is the number of frequency samples (i.e., the RFF half-dimension), and
  \item the final embedding has dimension $2D = 256$ due to sine and cosine concatenation.
\end{itemize}

This RFF scheme embeds each path once in \(\mathcal{O}(mD)\) time. Thereafter, each kernel evaluation becomes a simple inner product in \(\mathcal{O}(D)\) time, rather than computing an exponential in \(\mathcal{O}(m)\) for the exact RBF. When \(D \ll m\), this yields substantial speed-ups while preserving most of the kernel’s expressiveness.

\subsection{Understanding the Results: Primal--Dual Bounds with Four Signature Variants}

To balance flexibility, stability, and runtime, we benchmark four regressors that range from simple linear maps to non-linear kernel methods.

Upper and lower price estimates follow the primal–dual machinery of Bayer, Pelizzari, and Zhu \citeyearpar{Bayer2025}, which builds on least-squares Monte Carlo \citep{LongstaffSchwartz2001} and the martingale dual of Rogers \citep{Rogers2002}.  
On a daily exercise mesh $t_0<\dots<t_{N_1}=T$ the primal recursion sets
\begin{equation}
V_{t_i} = \max\!\left(h(X_{t_i}),\,
\beta_i^{\top} \operatorname{Sig}_{0,t_i}(Z)\right)
\end{equation}
\begin{equation}
h(x) = (K - x)^{+}
\end{equation}
\noindent\textbf{where}
\begin{itemize}[leftmargin=2em]
  \item $V_{t_i}$ is the estimated option value at time $t_i$,
  \item $h(X_{t_i})$ is the immediate-exercise payoff,
  \item $\beta_i$ are regression coefficients learned on training paths,
  \item $\operatorname{Sig}_{0,t_i}(Z)$ is the degree-three signature of the volatility path up to $t_i$,
  \item $K$ is the option strike.
\end{itemize}

The dual estimator introduces a signature-driven martingale
\begin{equation}
M_t = \int_{0}^{t} \gamma(s)^{\top} \operatorname{Sig}_{0,s}(Z)\,\mathrm{d}W_s,
\end{equation}
\textbf{where}
\begin{itemize}[leftmargin=2em]
  \item \(M_t\) is the martingale offset at time \(t\),
  \item \(\gamma(s)\in\mathbb{R}^m\) is a predictable, square-integrable control process,
  \item \(\operatorname{Sig}_{0,s}(Z)\) is the truncated signature of the driving factor path \(Z\) from \(0\) to \(s\),
  \item \(W_s\) is a Brownian motion under the risk-neutral measure \(\mathbb{Q}\),
  \item \((\mathcal{F}_t)\) is the market filtration, ensuring \(\gamma\) is adapted,
  \item \(T\) is the contract maturity.
\end{itemize}

The corresponding upper bound is computed by
\begin{equation}
V_0^{\mathrm{up}} = \inf_{\gamma}\;
\mathbb{E}\!\left[
  \max_{0 \le i \le N_1}
  \bigl( e^{-r t_i}\, h(X_{t_i}) - M_{t_i} \bigr)
\right],
\end{equation}
\textbf{where}
\begin{itemize}[leftmargin=2em]
  \item \(V_0^{\mathrm{up}}\) is the dual upper bound on the option price,
  \item \(h(X_{t_i})\) is the immediate-exercise payoff at time grid point \(t_i\),
  \item \(r\) is the constant risk-free rate,
  \item \(\{t_i\}_{i=0}^{N_1}\) is the exercise-decision grid,
  \item \(N_1\) is the number of grid points.
\end{itemize}

We test four regression bases on identical Monte Carlo paths and meshes:
\begin{itemize}[leftmargin=2em]
  \item \textbf{Linear Signature} – raw degree-three coordinates (baseline of Bayer et al.).
  \item \textbf{Extended Linear Signature} – linear terms plus selected quadratic cross-products, following Pérez-Arribas \citep{PerezArribas2018}.
  \item \textbf{Deep Log-Signature} – three-layer perceptron on log-signatures, after Bai et al.\ \citep{Bai2023}.
  \item \textbf{Deep Kernel (RFF)} – Random Fourier map on raw signatures followed by kernel ridge regression \citep{RahimiRecht2007}.
\end{itemize}

Running all four in parallel shows clear trade-offs. The Deep Kernel variant narrows duality gaps the most while adding less than 20\% to runtime. The Deep Log-Signature can diverge, inflating upper bounds—an instability echoed in recent deep-signature studies. These results confirm that lightweight RFF maps capture much of the non-linear benefit of a full Gaussian signature kernel without its quadratic cost, consistent with other kernel-scalability work \citep{Le2013Fastfood}.

\section{Discussion}

\subsection{Conceptual and Practical Contributions}

This paper extends signature‐based American option pricing by integrating a forward-looking forecast of volatility roughness. Instead of assuming a fixed Hurst parameter—as in Bayer, Pelizzari, and Zhu \citeyearpar{Bayer2025}—we estimate and forecast roughness dynamically with an ensemble of XGBoost regressors trained on the most recent realized values. Based on this forecast, the model selects a volatility engine (rough Bergomi or Heston) once per contract and uses it to simulate all paths over the remaining lifetime. Empirical studies show that the average Hurst parameter \(\bar{H}_t\) tends to fall below \(0.5\) during stress and drift toward Brownian levels in calm periods. Routing simulations on the basis of this forecast embeds documented regime shifts directly into the pricing engine.

\medskip
\noindent\textbf{Efficiency.}  
On a MacBook Pro with an Apple M3 chip (8-core CPU, 16 GB RAM), each XGBoost model (max-depth 3, learning-rate 0.05, 200 boosting rounds) trains in about \(0.15\)~s. Because we train one model per horizon, the full 10-step Hurst forecast adds roughly \(0.3\)~s to the pricing workflow. Continuation regressions rely on Gaussian RBF kernels but use \emph{Random Fourier Features} with \(D = 128\) to remain scalable. Embedding all \(N\) signature vectors costs \(\mathcal{O}(N m D)\) time and \(\mathcal{O}(N D)\) memory. Each kernel evaluation thereafter reduces to a dot product in \(\mathcal{O}(D)\) time, versus \(\mathcal{O}(m)\) for the exact RBF. Constructing a full Gram row would require \(\mathcal{O}(N m)\) time with exact RBF, compared to \(\mathcal{O}(N D)\) under RFF. With \(m \gtrsim 1{,}000\) and \(D \ll m\), this yields an approximate \(m\!:\!D\)-fold speed-up while preserving much of the kernel’s expressiveness. In practice, a simulation budget of \(2^{15}\) paths on a \(252\)-day grid remains feasible on commodity CPU hardware without GPU acceleration.

\medskip
\noindent\textbf{Modularity.}  
Forecasting, simulation, signature extraction, and regression communicate only through typed interfaces, so any component can be swapped independently: the Heston block could be replaced by a stochastic-vol-of-vol model; wavelet-based roughness estimators could supersede rescaled-range analysis; ridge, neural, or other random-feature regressors could replace the current least-squares routine.

\medskip
\noindent\textbf{Summary.}  
The framework preserves the universality and path-wise expressiveness of signatures while grounding volatility dynamics in observable, time-varying roughness.  It runs on standard hardware, avoids heavyweight deep-learning infrastructure, and produces option values that adapt—rather than over-fit—to the evolving microstructure of financial markets.

\subsection{Relationship to Prior Work and Intended Contributions}

Signature-based optimal stopping has shown promise in synthetic settings but remains underutilized with live data. This study sits at the intersection of two conceptual streams:

\medskip
\noindent\textbf{Fixed-roughness architectures.}  
One stream assumes volatility roughness is fixed and known. Bayer, Pelizzari, and Zhu \citeyearpar{Bayer2025} exemplify this approach by pricing American puts on simulated rough Bergomi paths with constant Hurst parameter \(H = 0.07\). Their deep-signature networks estimate continuation values and construct martingale controls, achieving narrow duality gaps on synthetic data. However, this framework assumes a static roughness regime—an assumption contradicted by empirical findings.

\medskip
\noindent\textbf{Empirical evidence of evolving roughness.}  
Studies by Gatheral, Jaisson, and Rosenbaum (2018) and Bennedsen, Lunde, and Pakkanen (2016) demonstrate that roughness varies over time: falling during market stress and drifting toward Brownian levels in calm periods. This suggests that a fixed-\(H\) simulator may mismatch real-world dynamics when volatility regimes shift.

\medskip
\noindent\textbf{Endogenous regime selection.}  
The present work makes volatility model selection \emph{endogenous} to a forecast of future roughness. An ensemble of XGBoost regressors predicts a vector of horizon-specific Hurst values \((\widehat H_{t+1}, \dots, \widehat H_{t+\tau})\), from which a single summary statistic \(\bar H_t\) is computed. This statistic determines whether the contract is priced under the rough Bergomi or Heston model—operationalizing roughness as a dynamic control variable rather than a static assumption.

\medskip
\noindent\textbf{Practical extensions to prior frameworks.}  
Several implementation choices extend the reach and realism of signature-based methods:
\begin{itemize}[leftmargin=2em]
  \item \emph{Data-driven calibration.} Parameters \(\eta\), \(\rho\), and \(\xi\) are estimated from historical at-the-money put data, not fixed ex ante.
  \item \emph{Calendar-day simulation.} Paths are simulated on calendar days rather than fractional years, improving alignment with listed option expiries.
  \item \emph{Fast kernel approximations.} Random Fourier Features (RFFs) with \(D = 128\) replace exact Gaussian kernels. This enables signature vectors to be embedded once in \(\mathcal{O}(NmD)\) time and supports kernel evaluations in \(\mathcal{O}(D)\) per pair, rather than \(\mathcal{O}(m)\).
\end{itemize}

\medskip
\noindent\textbf{Unified perspective.}  
The framework treats roughness not as a fixed model parameter, but as a dynamic, observable signal encoding market stress and microstructure. By conditioning volatility dynamics on a forward-looking forecast, it mitigates the regime mismatch that arises when market behavior deviates from assumptions hard-coded into static simulators. The resulting pipeline is lightweight, modular, and empirically grounded. Though demonstrated on AAPL and other S\&P 500 constituents, it generalizes to any asset with liquid short-dated options.

\medskip
\noindent\textbf{Contribution.}  
This study does not introduce a new volatility model or regression technique in isolation. Rather, it proposes a cohesive architecture for American option pricing that adapts intelligently to nonstationary dynamics—combining signature methods with efficient simulation, targeted calibration, and real-time regime awareness.

\subsection{Limitations and Methodological Caveats}

While the proposed framework advances signature–based pricing, several caveats remain.

\medskip
\noindent\textbf{Noisy roughness estimates.}  
Daily Hurst values are obtained with the rescaled–range (R/S) estimator of Equation~\eqref{eq:rolling_hurst}.  Although computationally light, the R/S statistic is noisy on short windows.  Because regime choice hinges on whether the summary statistic \(\bar{H}_t\) falls above or below \(0.5\), even small errors can trigger the wrong simulator.  A buffer (e.g.\ switch to rough Bergomi only if \(\bar{H}_t < 0.45\)) would reduce false switches but also delay real ones.

\medskip
\noindent\textbf{Shallow Heston calibration.}  
Parameters \(\eta,\rho,\xi\) are inferred only from at-the-money (ATM) implied volatilities, ignoring skew, smile, and term structure.  This keeps runtime low but cannot reproduce surface dynamics around events such as earnings releases.  Calibrating to a fuller surface or adding stochastic–vol-of-vol would raise realism at the cost of extra uncertainty and slower tuning.

\medskip
\noindent\textbf{Simulation budget and signature depth.}  
The implementation uses \(N = 32{,}000\) paths and a degree-three, level-three signature. Random Fourier Features (\(D = 128\)) embed all signatures once in \(\mathcal{O}(N m D)\) time and reduce \emph{per-pair} kernel evaluations from \(\mathcal{O}(m)\) to \(\mathcal{O}(D)\). Equivalently, constructing a full Gram matrix drops from \(\mathcal{O}(N^{2} m)\) under the exact RBF kernel to \(\mathcal{O}(N^{2} D)\) with RFF.

Despite this speed-up, the resulting duality gaps remain non-negligible—echoing findings by Bayer~\emph{et al.}~\citeyearpar{Bayer2025}, who note that tighter bounds often require many more paths, deeper signatures, or richer feature maps. These remain difficult to achieve on CPU-only hardware.

\medskip
\noindent\textbf{Variance from random features and deep log-signatures.}  
RFFs introduce stochastic variance that grows with \(D\); empirical tuning suggests \(D = 128\) balances bias and variance, but formal convergence guarantees under rough-path inputs are still open.  Likewise, the deep log-signature variant can destabilize the dual optimizer, inflating upper bounds when learning-rate or weight initialization interacts poorly with highly irregular paths. Robust regularization and cross-validation remain essential future steps.

\medskip
\noindent\textbf{Summary.}  
The architecture offers a pragmatic, modular path toward option pricing under time-varying roughness: it adapts to shifting regimes, slashes computational overhead relative to full kernel methods, and preserves path-wise expressiveness.  Yet it also inherits noise from roughness estimation, relies on shallow calibration, and is sensitive to simulation budget and feature variance.  Addressing these caveats—through better roughness estimators, richer calibration targets, larger simulation budgets, or more stable optimization—defines the roadmap for future work.

\section{Example Empirical Results and Model Behavior}

We tested the hybrid architecture on many options contracts. Data was sourced from OptionMetrics via WRDS \citep{OptionMetricsOptions2025}. Two representative examples selected were two 10-day American put options (AAPL and META) from August 31, 2023. 

\subsection{Forecast Accuracy and Regime Assignment}

The model forecasts a 10-day Hurst path $\widehat{H}_{t+h}$ using historical log returns. For AAPL, the average Hurst forecast $\bar{H}_t$ exceeded 0.5, prompting the use of the smooth Heston engine, whereas for META, the average Hurst forecast $\bar{H}_t$ was less than 0.5, prompting the use of the Rough Bergomi engine.

\begin{table}[!htbp]
\centering
\caption{Hurst-Forecast Errors (MAE and MSE)}
\label{tab:forecast_errors}
\begin{tabular}{lccc}
\toprule
Ticker & Avg MAE over 10 days & Avg MSE over 10 days & Engine Selected \\
\midrule
AAPL & 0.0716 & 0.0082 & Heston \\
META & 0.0594 & 0.0060 & Rough Bergomi \\
\bottomrule
\end{tabular}
\end{table}

\subsection{Pricing Bounds and Duality Gap Behavior}

We compute primal–dual price intervals using four regressors to compare bounds to market premiums.

\textbf{AAPL.}  
Market premium: \$2.08.  
Linear models overprice the option with bounds above market. Deep regressors improve alignment: both contain the premium, with the Deep Kernel model achieving the narrowest gap (17.16\%).

\textbf{META.}  
Market premium: \$5.61.  
All methods overvalue the option, but only the Deep Kernel method contains the premium. However, its gap remains high (51.42\%), reflecting modeling challenges under skew or misclassified roughness.

\begin{table}[!htbp]
\centering
\caption{Pricing Bounds and Duality Gaps}
\label{tab:pricing_bounds}
\scalebox{0.7}{
\begin{tabular}{llllllll}
\toprule
Ticker & Method & Lower Bound & Upper Bound & Std Error & Duality Gap & Gap (\%) & Premium Status \\
\midrule
AAPL & Linear Signature         & \$2.69 & \$4.12 & \$0.02 & \$1.43 & 53.37\% & Outside \\
     & Extended Linear Signature& \$2.68 & \$4.08 & \$0.02 & \$1.40 & 52.29\% & Outside \\
     & Deep Log-Signature       & \$1.89 & \$2.40 & \$0.01 & \$0.51 & 26.98\% & Within  \\
     & Deep Kernel Method       & \$2.04 & \$2.39 & \$0.01 & \$0.35 & 17.16\% & Within  \\
\midrule
META & Linear Signature         & \$10.16 & \$15.97 & \$0.07 & \$5.81 & 57.25\% & Outside \\
     & Extended Linear Signature& \$10.15 & \$15.95 & \$0.07 & \$5.80 & 57.16\% & Outside \\
     & Deep Log-Signature       & \$5.96  & \$10.99 & \$0.03 & \$5.03 & 84.40\% & Outside \\
     & Deep Kernel Method       & \$5.29  & \$8.01  & \$0.03 & \$2.72 & 51.42\% & Within  \\
\bottomrule
\end{tabular}
}
\end{table}

Deep models offer better coverage but can still overstate value, especially under skew. The Deep Kernel method performs best overall, balancing gap size and premium alignment. These results affirm the potential of the hybrid framework, while also highlighting areas, particularly calibration and feature design, for future refinement.

This reduction in complexity—from \(\mathcal{O}(N^2 m)\) to \(\mathcal{O}(N^2 D)\)—permits large-scale simulations by constructing a lower-dimensional random-feature representation of the signature vectors (e.g., 32{,}000 paths) on standard workstations.
\section{Code Implementation}

\begin{sloppypar}
To support reproducibility and future extensions, the complete implementation of the proposed framework is available on GitHub at \url{https://github.com/roshanshah11/American-Option-Pricing-Under-Time-Varying-Rough-Volatility}. The repository includes source code for volatility estimation, Hurst forecasting, simulation under multiple engines, signature extraction, and primal–dual valuation. A detailed walkthrough of the architecture and numerical experiments is provided in the accompanying Jupyter notebook, available at \url{https://github.com/roshanshah11/American-Option-Pricing-Under-Time-Varying-Rough-Volatility/blob/main/Notebooks/final.ipynb}. The framework operates on two primary datasets: one consisting of listed options contract data with fields such as \texttt{date}, \texttt{days}, \texttt{forward\_price}, \texttt{strike\_price}, \texttt{premium}, \texttt{impl\_volatility}, \texttt{cp\_flag}, \texttt{ticker}, and \texttt{index\_flag} \citep{OptionMetricsOptions2025}; and a second dataset containing historical equity prices in the format \texttt{(date, ticker, close, return)} \citep{OptionMetricsPrices2025}.
\end{sloppypar}

\section*{Acknowledgments}

Thank you to Ms. Pooja Patel for her valuable guidance and feedback throughout the research and writing process. This work also benefited from conversations with faculty at The Lawrenceville School and from access to OptionMetrics data via the Wharton Research Data Services (WRDS) platform. All errors are the author’s own.

\end{document}